# Polarization dependent femtosecond laser modification of MBE-grown III-V nanostructures on silicon

Sander R. Zandbergen,[1] Ricky Gibson,[1,2] Babak Amirsolaimani,[1] Soroush Mehravar,[1] Patrick Keiffer,[1] Ali Azarm,[1] and Khanh Kieu[1]

[1]*College of Optical Sciences, University of Arizona, 1630 E University Boulevard., Tucson, Arizona 85721, USA*
[2]*Currently with the Air Force Research Laboratory, Sensors Directorate, 2241 Avionics Circle, Building 600, Wright Patterson Air Force Base, Ohio 45433, USA & University of Dayton Research Institute, 300 College Park, Dayton Ohio 45469-0101, USA*
*\*Corresponding author: szandbergen@optics.arizona.edu*

**Abstract:** We report a novel, polarization dependent, femtosecond laser-induced modification of surface nanostructures of indium, gallium, and arsenic grown on silicon via molecular beam epitaxy, yielding shape control from linear and circular polarization of laser excitation. Linear polarization causes an elongation effect, beyond the dimensions of the unexposed nanostructures, ranging from 88 nm to over 1 µm, and circular polarization causes the nanostructures to flatten out or form loops of material, to diameters of approximately 195 nm. During excitation, it is also observed that the generated second and third harmonic signals from the substrate and surface nanostructures increase with exposure time.

## 1. Introduction

Since the development of ultrafast laser sources, especially the generation of femtosecond (fs) pulses from solid-state Ti:sapphire lasers in the early 1990s [1], there have been many studies of the interaction of ultrashort laser pulses with materials. This area has reached an advanced state of research, where many applications have been and are currently being developed, specifically in the area of fs laser material processing due to the minimum thermal and mechanical damage caused by fs lasers. A number of books and review articles have been written on this subject [2-5]. The main takeaway message of the literature is that these lasers not only allow for the study of the dynamics of fundamental processes in nature that involve changes in the structure of matter which occur on ultrashort timescales, but also that laser irradiation can be used to induce changes in the chemistry, crystal structure, and morphology in a variety of material systems. These systems comprise, for instance, transparent dielectrics such as glass [6], semiconductors such as silicon [7] and gallium arsenide (GaAs) [8,9], and biological specimens [10]. High-intensity fs lasers also provide an easy way of circumventing the diffraction limit by virtue of multiphoton absorption [11].

Much of the work cited above relies on the physical process of laser ablation, the process of removing material from a solid surface by irradiating it with an intense laser beam above a material-dependent threshold. Other studies have looked at plasmon-mediated ablation below the substrate ablation threshold using the near-field enhancement by gold nanoparticles [12-14]. Many researchers have even taken advantage of certain materials' propensity for reversible transitions/switching between stable amorphous and crystalline phases under various laser excitation [15-17], and these phase change properties have been utilized for many years in optical data storage. Particularly interesting is the formation of laser-induced periodic surface structures (LIPSS) which are ripples in the substrate surface due to fs laser irradiation [7,18,19]. This is widely accepted as a multiple-pulse, interference effect resulting in substrate modification above and below the ablation threshold.

Recently, it was determined that these morphological changes can occur without ablating or melting [20]. However, aside from that work, little research has gone into non-ablative material modification (NMM), where material is modified but not removed.

In this work, we demonstrate a non-ablative, polarization-sensitive control over nanostructure material movement with a focused fs laser beam. This NMM effect presents a new regime that has not been systematically studied. The laser parameters fall between probing without substrate modification and ablating the sample; this is a 'modification regime' where there does not appear to be mass removal occurring, but one in which the nanostructures change in shape. While we do not possess the means to monitor the mass during exposure, we qualitatively estimated the volumes before and after exposure, deducing that approximately the same amount of material remains post-exposure. Even with conservative estimates, our laser parameters are well below the ablation damage thresholds of silicon (200 mJ/cm2) [7] and GaAs (175 mJ/cm2) [8], evidenced by the integrity of the crystallinity of the samples following laser exposure. Our laser parameters are near the ablation threshold of indium (2.5 mJ/cm2), as determined by Götz and Stuke using 500 fs UV laser excitation (248 nm) [21]. If the material movement that we observe arises solely from interactions with the ultrafast field, then the material movement may be much faster than the speed of sound in it. Several experimental parameters have been identified to be important for the modification and control of these nanostructures. These parameters are the laser wavelength, laser beam polarization, the laser fluence (energy density, J/cm2) or intensity (W/cm2), and the number of laser pulses N applied to the same spot. While we have not yet fully established the physical mechanism of this effect, we speculate that it can be ported to other material systems and therefore result in a rich vein of new research.

## 2. Sample growth

The sample in this work was grown by solid source molecular beam epitaxy (MBE) on an n-doped silicon (100) substrate cut on-axis. The 2-inch silicon substrate was prepared for growth with a standard hydrofluoric acid (48%) dip for approximately 3 minutes to remove the native oxide layer and hydrogen passivate the surface and then rinsed with DI water. The hydrogen-terminated wafer was then quickly transferred to the MBE load chamber to prevent re-oxidation of the substrate surface. In the load chamber, the substrate and molybdenum substrate holder were outgassed at 200° C for more than 4 hours. The growth procedure followed similarly to that of one employed by Urbańczyk, et al. [22] who reported the growth of crystalline indium islands on GaAs (100) at low temperatures and their subsequent conversion into InAs islands/QDs under As flux for photoluminescence studies. We attempted to emulate this growth on silicon (100). First, 16 monolayers of indium were deposited at a substrate temperature of 200° C according to our Type C manipulator thermocouple. Our deposition rates are calibrated for the planar growth of InAs; the atomic flux used for indium corresponds to an InAs growth rate of 0.025 ML/s. Stranski-Krastanov (SK) growth was assumed to have taken place, where the initially deposited indium nucleates and grows two-dimensionally to form an epitaxial thin film, before reaching a critical layer thickness (between 2 and 3 ML), after which three-dimensional islands are formed [23]. Second, arsenic (As4 from a valved source) was deposited at a rate of 0.286 ML/s for just over one minute and then closed to allow the crystallization of indium and arsenic, in the aim of InAs island formation. Third, while the crystallization is taking place, the substrate temperature was increased to approximately 500° C over 7 minutes to allow for improved GaAs growth. Finally, after the substrate temperature had mostly stabilized, 5 nm of GaAs were grown with a rate of 0.191 ML/s to cap the expected InAs QDs for quantum-confined emission and

PL studies. The growth was ended by closing all mechanical shutters in the chamber and cutting off the substrate heater current. After the substrate temperature dropped below 400° C, the substrate was removed from the epitaxy chamber.

The growth produced unique nanostructures consisting of indium, gallium, and arsenic, as seen in Fig. 1. They are consistently longer in one direction that aligns with the crystal axes of the (100) silicon substrate (inset Fig. 1(a)). There is semiconductor material, in the form of small islands, distributed across the sample between the larger nanostructures, indicating that any thin film that formed via SK growth at the onset of indium deposition broke up when the arsenic and gallium were deposited. Looking more closely at the nanostructures in Fig. 1(b) and (c), it is clear that there are two parts of the nanostructure; one that is rounded and one that is faceted.

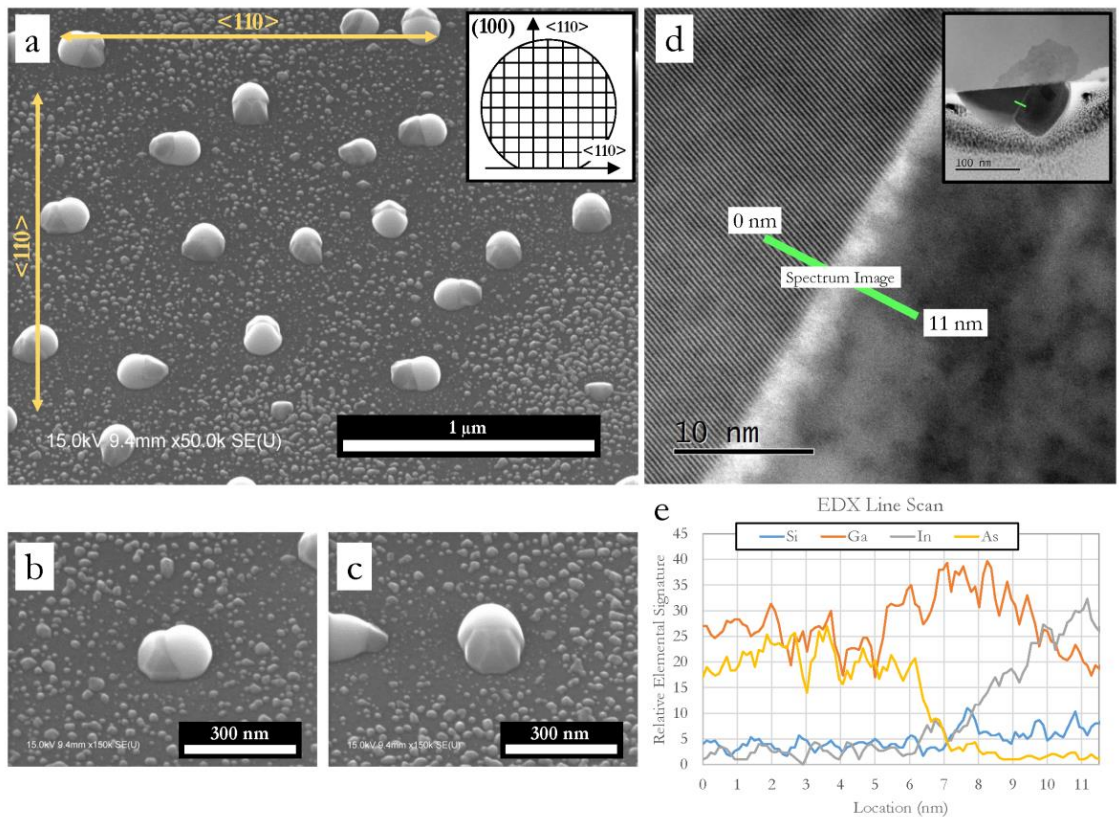

Fig. 1. SEM, TEM, and EDX analysis of nanostructures. SEM images (45°) (a) through (c) of the unexposed semiconductor sample with III-V nanostructures grown via MBE on n-doped silicon (100) substrate. In (a) the nanostructures have directionality and point either up/down or left/right, aligning to the crystal axes of the silicon substrate (inset), indicated by the yellow arrows. In the two close-ups (b) and (c), the individual structures are half-rounded and half-faceted. (d) A high-resolution TEM image zoomed into the interface between the two regions shown in the inset low-magnification TEM image. An EDX line scan (e) was taken between these two regions along the green line labeled "Spectrum Image". Using the EDX data, we can determine that the left side (faceted) of the interface is a crystalline GaAs by the strong lattice lines, and the right side (rounded) of the image is predominantly indium.

To better understand the atomic structure and elemental composition of these nanostructures, an initial investigation was completed via transmission electron microscopy (TEM) analysis. An FEI Helios Nanolab 660 DualBeam Microscope focused ion beam (FIB) device was used to manufacture an ultra-thin lamella for cross-sectional S/TEM. Two different aberration-corrected S/TEMs were employed to probe the lamella, a JEOL ARM200F whose detectors allow for high-angle annular dark field (HAADF) imaging and a FEI Titan 80-300.

The initial TEM data verify that the as-grown nanostructures consist of two regions, as suggested by scanning electron microscope (SEM) imaging. One part is a crystalline GaAs, and the other part is a mixture of indium and gallium. Fig. 1(d) shows a high-resolution TEM (HR-TEM) image which zooms in on the interface between the two regions of a nanostructure (inset Fig. 1(d)). Clear lattice lines on the left of the TEM image demonstrate the crystallinity of that region which corresponds to the energy dispersive x-ray spectroscopy (EDX) line scan in Fig. 1(e). EDX is a spectroscopic technique used to determine a sample's elemental composition. The EDX line scan in Fig. 1(e) shows the crystalline region to be GaAs. There appears to be a layer of gallium separating the two regions. The region on the right is an indium-heavy mixture of indium and gallium, but as you move farther into the right side, the indium to gallium ratio increases even further, suggesting the mixture is actually predominantly indium. This indium region appears to be largely amorphous. We hypothesize that as the substrate temperature is increased, the deposited crystalline indium melts and reforms as the observed rounded regions. For this report, we study the behavior of these half-crystalline nanostructures of indium, gallium, and arsenic under intense fs laser irradiation.

## 3. Irradiation via multiphoton microscopy

To characterize and determine if these nanostructures exhibited strong nonlinear response, the samples were exposed to a focused fs laser beam in a custom-built multiphoton microscope (MPM), previously reported by Kieu, et al. [24]. Multiphoton microscopy is a laser-point scanning imaging technique where the nonlinear signals act as the contrast mechanism. The main application of this technique is in biomedical imaging [10, 25], but it can also be used to characterize nonlinear optical properties of materials [26, 27].

The light source in the system is an amplified erbium-doped mode-locked fiber laser outputting linearly polarized light centered at 1560 nm. The maximum average power of the laser is ~50 mW with a repetition rate of ~8 MHz and ~150 fs pulse duration. Therefore, the pulse peak power is estimated to be ~41 kW, and the pulse energy is ~6.25 nJ. The laser beam is scanned with a 2D galvo mirror system, yielding a field of view (FOV) of approximately 300x300 µm and focused on the sample using a 20x microscope objective (New Focus 5724-C-H) with a numerical aperture (NA) of 0.5. This NA allows for maintaining the polarization state as the light is focused onto the sample. Assuming linear polarization, NA near or above 1 will have a non-negligible z-component [28]. The measured spot size is near the linear diffraction limit (1.22λ/2NA) of ~2.0 µm, giving a calculated intensity of ~0.25 TW/cm2 which should lead to a strong nonlinear response from the sample. The backscattered signals of second and third harmonic generation (SHG and THG) light from each point on the sample are separated into two paths using a long-pass dichroic mirror (cut-off at 562 nm) and then detected using photomultiplier tubes (Hamamatsu H8249 and H7732). Narrow band-pass filters are used to select SHG (FF01-780/12-25) and THG (FF01-517/20-25) signals at central wavelengths of 780 nm and 520 nm, respectively. See the schematic in Fig. 2.

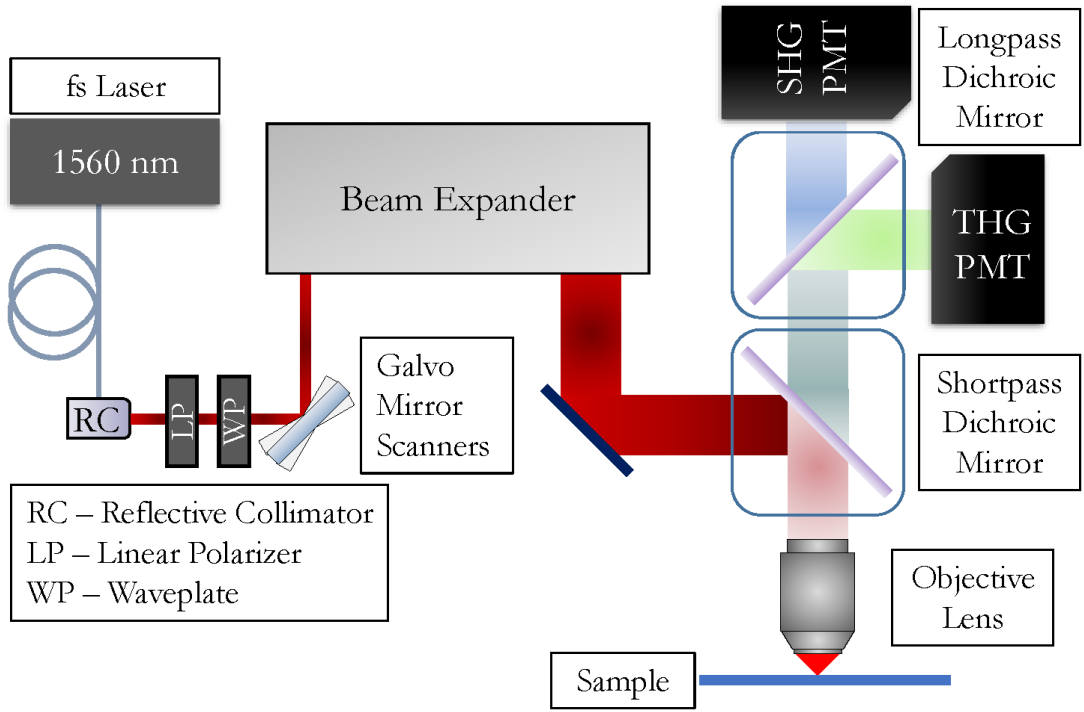

Fig. 2. Schematic diagram of multiphoton microscope. A linearly polarized laser beam from a mode-locked fiber laser at 1560 nm is first collimated and passes through a linear polarizer (for power control) and wave plate (for polarization control). The laser light is then scanned with the galvo mirrors, expanded with the all-reflective beam expander, reflected off the shortpass dichroic mirror, and directed through the objective lens, focusing tightly onto the sample. The reflected and emitted light travels back up through the lens and shortpass dichroic mirror and enters the SHG and THG photomultiplier tubes (PMTs), after passing through or reflecting off of a longpass dichroic mirror. Narrow bandpass filters are used to guarantee the signal is from SHG, THG, or fluorescence.

## 4. Observation of laser-induced modification

As previously discussed, the high-intensity fs laser beam is focused and raster-scanned across the sample with a FOV of about 300x300 µm. Strong nonlinear signals were generated, and at the stated intensity level, we observed that the nonlinear (SHG and THG) signals increased with exposure time as shown in Fig. 3. Qualitatively, there is a significant increase in signal from the first to the last exposure. To quantify this changing signal, the PMT pixel values were averaged using the open-source software Fiji [29] and plotted versus exposure number. The resulting plot demonstrates the increase in SHG and THG signal with each exposure by the fs laser, suggesting a modification of the sample via change in the shape, size, composition, and/or crystal structure of the nanostructures.

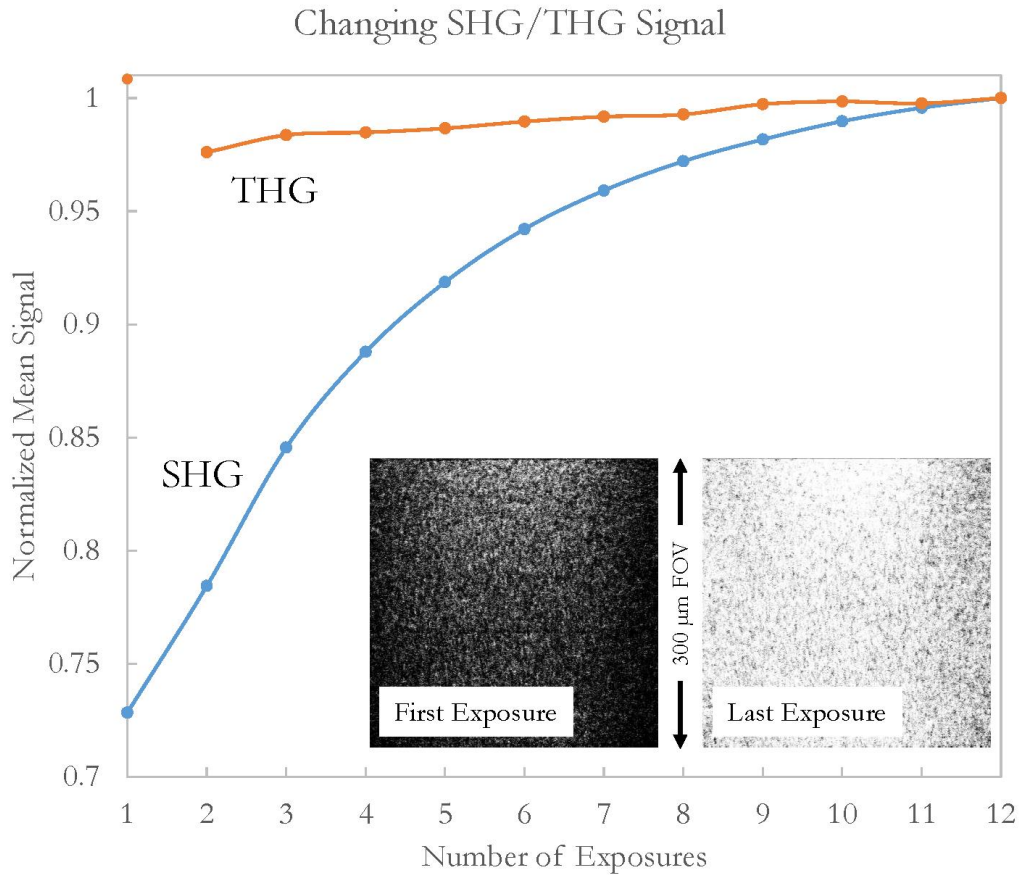

Fig. 3. The changing SHG and THG signal as a function of the number of exposures. The inset images show the first and final SHG PMT output. The FOV of the image is approximately 300x300 µm. The plot clearly shows the increase in SHG and THG signal with each exposure by the linearly polarized fs laser. An increase in signal was also observed during exposure with circular polarization. The optical resolution of the system is not high enough to resolve individual nanostructures.

Scanning electron microscopy was used to investigate the physical shape of the nanostructures following fs laser exposure and to initially verify that indeed the nanostructures on the surface had changed shape. The initial results showed that the exposed nanostructures became elongated. To identify that this elongation was due to the linearly polarized laser light, a series of samples were then exposed with various angles of linear polarization, utilizing a half-wave plate, and compared with circular polarization, utilizing a quarter-wave plate. The SEM was also used to characterize the results of these different polarization tests, with results shown in Fig. 4. We observed a clear laser-induced change in shape, which follows the laser's polarization direction, along with an increasing nonlinear signal. For circular polarization, the material is deformed into a circular shape, confirming that the polarization has a significant impact on the final shape of the modified nanostructures. In fact, the circular material movement direction depends on polarization handedness, shown by the red circular arrows in Fig. 4(c). Every incomplete circle follows a clockwise (CW) path from the GaAs region. Upon irradiation with the opposite circular polarization, the material follows a CCW circular path (not shown). This is believed to be the first reported observation of a non-ablative, polarization-sensitive control over surface nanostructure material movement with a high-intensity fs laser.

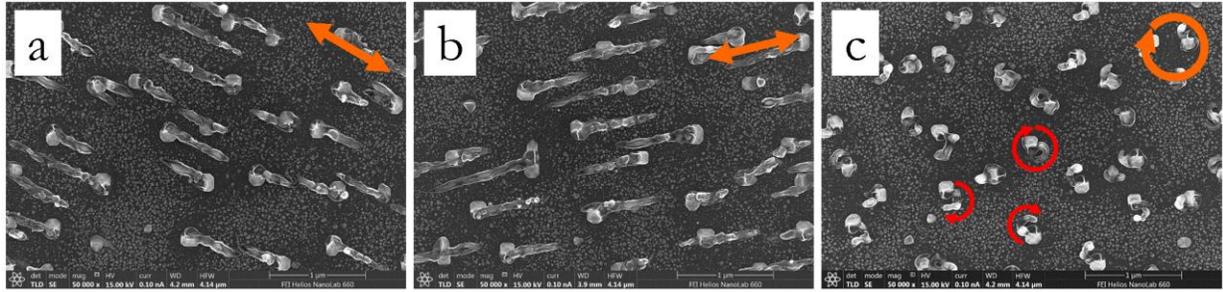

Fig. 4. Observation of non-ablative material modification (NMM). SEM images clearly demonstrate a laser-induced change in shape, along the direction of laser excitation polarization, which resulted in increasing nonlinear signal. In (a) and (b), the angle of linear polarization is indicated by the orange arrows, and in (c), circular polarization is indicated by the circle. The red circular arrows in (c) indicate the CW material movement direction.

The elongation effect, measured beyond the original dimensions of the nanostructures, can be controlled by strength and duration of exposure, as seen in Fig. 5. A 4x4 array of areas were exposed for varying lengths of time at various powers. Output power was controlled by a linear polarizer (LPNIR100-MP2) and had a maximum value of approximately ~42 mW when aligned for maximum transmission of the linearly polarized laser light. Three lower powers (~16 mW, ~29 mW, and ~39 mW) were set by rotating the polarizer. The areas were exposed by the laser once, and the scan duration was controlled by changing the galvo scan speed. The frame rates, i.e. inverse scan times, in frames per second (fps) were approximately 0.488 fps, 0.244 fps, 0.122 fps, and 0.076 fps. The associated pixel dwell times were, respectively, 1.95 µs, 3.91 µs, 7.81 µs, and 12.5 µs. A pulse picker was not used to directly control the number of pulses impinging on the sample, but with a repetition rate of 8 MHz, the stated pixel dwell times, and an approximation on the spatial overlap of the focused beam as it scans across the FOV, we estimated the number of pulses to be about 160, 320, 640, and 1024. Fig. 5(a) is an SEM image of a small part of the area exposed once at 0.488 fps at the highest power. The average length of the elongation effect was $150 \pm 22$ nm. Fig. 5(b) is an SEM image of a small part of the area exposed once at 0.076 fps at the highest power. The average length of the elongation effect was $249 \pm 57$ nm. Fig. 5(c) plots the rest of the data from the 4x4 exposed array. The length data were acquired using Fiji [29], and almost 750 elongated nanostructures were measured and averaged. The error bars are the standard deviation of these measurements. At each power, as the exposure time (number of pulses) is increased, the nanostructures are further elongated. Also, as power is increased to about 40 mW, there is again an increase in the length of these nanostructures. By exposing areas more than once, as in Fig. 4(a) and (b), elongation lengths of up to and over 1 µm can be reached. We also measured the effect of circularly polarized light, resulting in the nanostructures of Fig. 4(c); the induced circular shapes have outer diameters of $195 \pm 16$ nm, with thicknesses of $66 \pm 5$ nm (25 islands analyzed).

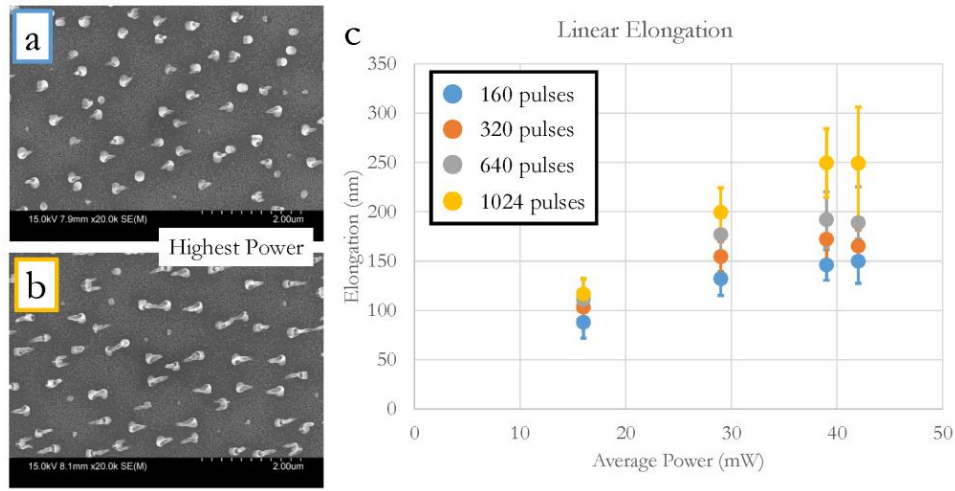

Fig. 5. Power and pulse number dependence on elongation effect. (a) SEM image of the area exposed with the least number of pulses at the highest power. (b) SEM image of the area exposed with the largest number of pulses at the highest power. (c) As you increase the number of pulses and/or the power of your irradiation, there are clear trends toward a stronger elongation effect.

In an attempt to better understand the physical mechanism behind this observed effect, another TEM sample was prepared by FIB for further atomic structure and elemental analysis. From inspection of the SEM images of exposed nanostructures, it appears that the rounded, indium side of the nanostructure is moving, while the crystalline, GaAs side remains stationary. The STEM image in Fig. 6(a) supports this observation. It appears that the indium is energetically propelled along the direction of polarization and even enters into the silicon substrate, and the GaAs side remains intact and unmoved. The EDX line scans in Fig. 6(b) and (c) show the location of the elements in the image.

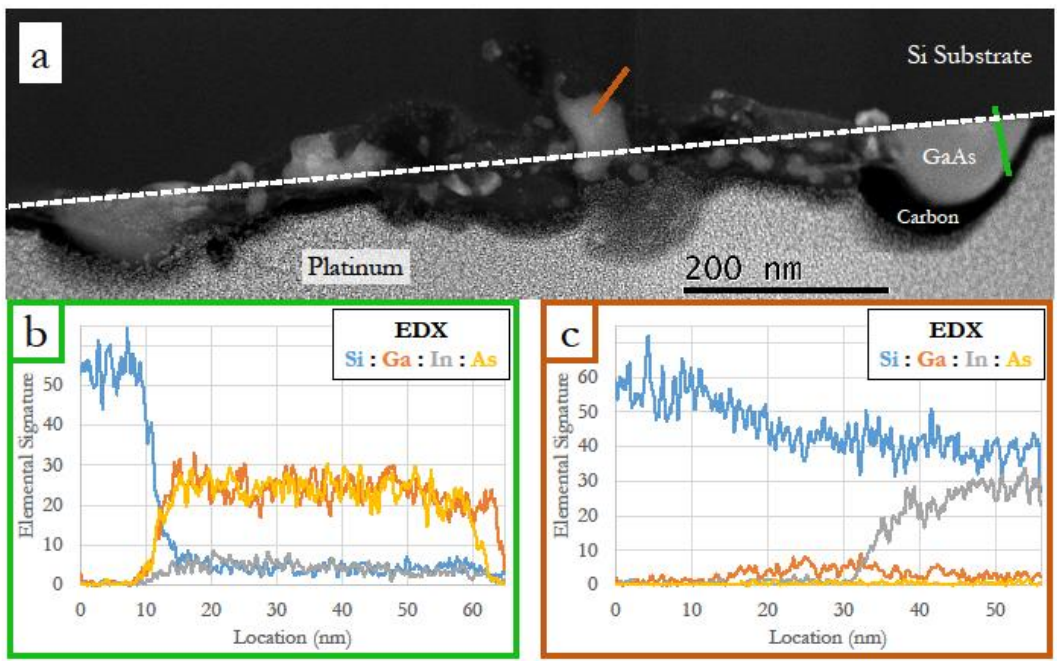

Fig. 6. TEM and EDX analysis of exposed nanostructures. A STEM image (a) of two nanostructures that have connected after being exposed. The silicon substrate surface (dotted line), the protective layer of carbon, and the protective layer of platinum are indicated. The EDX line scans (b,c) show the relative elemental signature versus location in nanometers along the lines indicated in the image.

## 5. Discussion and outlook

We are currently considering two mechanisms to explain this observed effect, that it arises thermally or that it arises from interactions with the strong electric field of the laser pulses. These two mechanisms have different time scales. A thermal effect has a time scale many orders of magnitude slower than what one would have with the laser field effect [6]. In the thermal realm, the high intensity laser field heats up the electron gas, and the absorbed energy is then transferred to the lattice after much more than a picosecond time delay (after the laser pulse has ended) and melts the material, but this should not be influenced by the polarization of the laser as was observed here. The duration of these laser pulses is about $\tau \sim 150$ fs, so if the effect is dominated by the strong laser field interaction rather than the thermal effect, then one may be observing material movement velocities much larger than the speed of sound in solids. A pump-probe apparatus is currently being built to investigate the time dynamics of this effect, and if it is confirmed, it may open the door to new applications, for example, where ultrafast pulses of light could be used to accelerate objects or material to very high velocities within an extremely short time. Femtosecond laser pulses have been investigated for nanostructuring or micromachining in a variety of material systems, including transparent and absorptive materials. Many of these past studies focused on the effect of short-pulse laser ablation to remove materials or to change a material's properties, but the current study is believed to be the first observation of a form of fs laser non-ablative nanostructure modification where the excitation polarization influences the material movement direction.

These observations demonstrate the possibility of after-growth modification of MBE nanostructures with fs laser pulses. This may also open the door to the modification of MBE growths, where one illuminates the growth interface with a fs laser through a viewport to optimize the growth process or to construct/fabricate nanostructures in situ. Lasers have been used in MBE as a method to heat sources [30], a technique known as laser assisted deposition and annealing, but they have not been used to investigate surface modification during growth. Chopra, in 1965, looked at the growth of thin films under an applied electric field [31], yet no one has investigated this process with a fs laser. One can also imagine this as a new paradigm of controllable phase-change or reconfigurable materials that take advantage of III-V material properties and plasmonics.

## Acknowledgments

We gratefully acknowledge funding by the Air Force Office of Scientific Research (AFOSR, Dr. Gernot Pomrenke) (FA9550-15-1-0389). This work utilized facilities within the LeRoy Eyring Center for Solid State Science at Arizona State University. We thank the late Prof. Galina Khitrova for all her support on this project. We thank Dr. A. Azarm for discussions about this research.## References

1. D. E. Spence, P. N. Kean, and W. Sibbett, "60-fsec pulse generation from a self-mode-locked Ti:sapphire laser," Opt. Lett. **16**, 42-44 (1991).

2. E. Gamaly, Femtosecond Laser-Matter Interactions: Theory, Experiments and Applications (Pan Stanford, 2011).

3. K. Sugioka and Y. Cheng, Ultrafast Laser Processing: From Micro- to Nanoscale (Pan Stanford, 2013).

4. E. G. Gamaly and A. V. Rode, "Physics of ultra-short laser interaction with matter: From phonon excitation to ultimate transformations," Prog. Quant. Electron **37**, 215-323 (2013).

5. K. C. Phillips, H. H. Gandhi, E. Mazur, and S. K. Sundaram, "Ultrafast laser processing of materials: a review," Adv. Opt. Photon. **7**, 684-712 (2015).

6. R. R. Gattass and E. Mazur, "Femtosecond laser micromachining in transparent materials," Nat. Photonics **2**, 219-225 (2008).

7. J. Bonse, S. Baudach, J. Krüger, W. Kautek, and M. Lenzner, "Femtosecond laser ablation of silicon-modification thresholds and morphology," Appl. Phys. A **74**, 19-25 (2002).

8. A. Cavalleri, K. Sokolowski-Tinten, J. Bialkowski, and D. van der Linde, "Femtosecond laser ablation of gallium arsenide investigated with time-of-flight mass spectroscopy," Appl. Phys. Lett. **72**, 2385 (1998).

9. S. Liu, M. B. Sinclair, S. Saravi, G. A. Keeler, Y. Yang, J. Reno, G. M. Peake, F. Setzpfandt, I. Staude, T. Pertsch, and I. Brener, "Resonantly Enhanced Second-Harmonic Generation Using III-V Semiconductor All-Dielectric Metasurfaces," Nano. Lett. **16**, 5426-5432 (2016).

10. W. Denk, J. H. Strickler, and W. W. Webb, "Two-Photon Laser Scanning Fluorescence Microscopy," Science **248**, 73-76 (1990).

11. S. Kawata, H. B. Sun, T. Tanaka, and K. Takada, "Finer features for functional microdevices," Nature **412**, 697–698 (2001).

12. A. Plech, V. Kotaidis, M. Lorenc, and J. Boneberg, "Femtosecond laser near-field from gold nanoparticles," Nat. Phys. **2**, 44-47 (2006).

13. N. N. Nedyalkov, T. Miyanishi, and M. Obara, "Enhanced near field mediated nanohole fabrication on silicon substrate by femtosecond laser pulse," Appl. Surf. Sci. **253**, 6558-6562 (2007).

14. A. Kolloch, P. Leiderer, S. Ibrahimkutty, D. Issenmann, and A. Plech, "Structural study of near-field ablation close to plasmon-resonant nanotriangles," J. Laser Appl. **24**, 042015 (2012).

15. C. N. Afonso, J. Solis, F. Catalina, C. Kalpouzos, "Ultrafast reversible phase change in GeSb films for erasable optical storage" Appl. Phys. Lett. **60**, 3123-3125 (1992).

16. B. Soares, F. Jonsson, and N. I. Zheludev, "All-Optical Phase-Change Memory in a Single Gallium Nanoparticle," Phys. Rev. Lett. **98**, 153905 (2007).

17. A. Karvounis, B. Gholipour, K. F. MacDonald, and N. I. Zheludev, "All-dielectric phase-change reconfigurable metasurface," Appl. Phys. Lett. **109**, 051103 (2016).

18. M. Birnbaum, "Semiconductor Surface Damage Produced by Ruby Lasers," J. Appl. Phys. **36**, 3688 (1965).

19. J. Bonse, J. Krüger, S. Höhm, and A. Rosenfeld, "Femtosecond laser-induced periodic surface structures," J. Laser Applications **24**, 042006 (2012).

20. M. J. Abere, B. Torralva, and S. M. Yalisove, "Periodic surface bifurcation induced by ultrafast laser generated point defect diffusion in GaAs," Appl. Phys. Lett. **108**, 153110 (2016).

21. T. Götz and M. Stuke, "Short-pulse UV laser ablation of solid and liquid metals: indium," Appl. Phys. A **64**, 539-543 (1997).

22. A. Urbańczyk, G. J. Hamhuis, and R. Nötzel, "In islands and their conversion to InAs quantum dots on GaAs (100): Structural and optical properties," J. Appl. Phys. **107**, 014312 (2010).
23. J. Knall, J.-E. Sundgren, G. V. Hansson, and J. E. Greene, "Indium overlayers on clean Si(100)2x1: Surface structure, nucleation, and growth," Surf. Sci. **166**, 512-538 (1986).
24. K. Kieu, S. Mehravar, R. Gowda, R. A. Norwood, and N. Peyghambarian, "Label-free multi-photon imaging using a compact femtosecond fiber laser mode-locked by carbon nanotube saturable absorber," Biomed. Opt. Express **4**, 2187-2195 (2013).
25. W. R. Zipfel, R. M. Williams, and W. W. Webb, "Nonlinear magic: multiphoton microscopy in the biosciences," Nat. Biotechnol. **21**, 1369-1377 (2003).
26. A. Säynätjoki, L. Karvonen, J. Riikonen, W. Kim, S. Mehravar, R. A. Norwood, N. Peyghambarian, H. Lipsanen, and K. Kieu, "Rapid Large-Area Multiphoton Microscopy for Characterization of Graphene," ACS Nano **7**, 8441-8446 (2013).
27. S. Shahin, S. Mehravar, P. Gangopadhyay, N. Peyghambarian, R. A. Norwood, and K. Kieu, "Multiphoton microscopy as a detection tool for photobleaching of EO materials," Opt. Express **22**, 30955-30962 (2014)
28. M. Mansuripur, Classical Optics and its Applications (Cambridge University, 2009).
29. J. Schindelin, I. Arganda-Carreras, and E. Frise, "Fiji: an open-source platform for biological-image analysis", Nat. Methods **9**, 676-682 (2012).
30. M. A. Herman and H. Sitter, Molecular Beam Epitaxy: Fundamentals and Current Status (Springer, 1996).
31. K. L. Chopra, "Growth of Thin Metal Films Under Applied Electric Field," Appl. Phys. Lett. 7, 140 (1965).